\documentclass[aps,prb,twocolumn,superscriptaddress,showpacs,floatfix]{revtex4}
\usepackage{graphicx}
\usepackage{epsfig}
\usepackage{amsmath}
\usepackage{amssymb}

\begin{document}

\title{Coupling of the Local Defect and Magnetic Structure of W\"ustite, Fe$_{1-x}$O}

\author{Paul J. Saines}
\email{paul.saines@chem.ox.ac.uk}
\affiliation{Department of Chemistry, University of Oxford, Inorganic Chemistry Laboratory, South Parks Road, Oxford OX1 3QR, U.K.}

\author{Matthew G. Tucker}
\affiliation{ISIS Facility, Rutherford Appleton Laboratory, Harwell Oxford, Didcot, OX11 0QX, U.K.}

\author{David A. Keen}
\affiliation{ISIS Facility, Rutherford Appleton Laboratory, Harwell Oxford, Didcot, OX11 0QX, U.K.}

\author{Anthony K. Cheetham}
\affiliation{Department of Materials Science and Metallurgy, University of Cambridge, Pembroke Street, Cambridge, CB2 3QZ, U.K.}

\author{Andrew L. Goodwin}
\affiliation{Department of Chemistry, University of Oxford, Inorganic Chemistry Laboratory, South Parks Road, Oxford OX1 3QR, U.K.}
\date{\today}
\begin{abstract}
The local nuclear and magnetic structure of w\"ustite, Fe$_{1-x}$O, and the coupling between them, has been examined using reverse Monte Carlo refinements of variable-temperature neutron total scattering data. The results from this analysis suggest that the individual units in a tetrahedral defect cluster are connected along $\langle110\rangle$ vectors into a Koch-Cohen-like arrangement, with the majority of octahedral vacancies concentrated near these defects. Bond valence calculations indicate a change in the charge distribution on the cations with the charge on the tetrahedral interstitials increasing on cooling. The magnetic structure is more complex than previously thought, corresponding to a non-collinear spin arrangement described by a superposition of a condensed spin wave on the established type-II antiferromagnetic ordering. This leads to an architecture with four groups of cations each with different spin directions. The cations within the interstitial clusters appear to be weakly ferromagnetically coupled and their spins are correlated to the spins of the octahedral cations closest to them. This work not only provides further insight into the local structure of w\"ustite but also a better understanding of the coupling between defect structures and magnetic and charge-ordering in complex materials.\end{abstract}

\pacs{61.05.F-,75.25.-j,61.66.Fn,61.72.Dd}

\maketitle

\section{Introduction}

Strong coupling between structural defects and magnetic, superconducting and charge ordering properties has been found in a wide range of important materials. The precise size and arrangements of atoms in a nano-cluster can control the magnetic ground state of a nanomagnet, tuning the strength of its magnetic interactions and inducing frustration.\cite{Khajetoorians_2012} Such coupling is also of great significance in the cuprate and iron pnictide high temperature superconductors.\cite{Orenstein_2000,Mazin_2008} The superconductivity of these compounds is strongly tied to the existence of antiferromagnetic spin-striped phases, whose formation and positions are tuned by cation and anion doping.\cite{Orenstein_2000,Mazin_2008} The well known Verwey charge-ordering transition in magnetite is also related to the formation of a complex trimeron structure, with these polarons fluctuating above the transition.\cite{Senn_2012} Despite these phenomena generating great interest, detailed crystallographic studies of coupling between the defect structure and the electronic properties they control remains difficult and have rarely been achieved. 

W\"ustite, Fe$_{1-x}$O, is a canonical example of such coupling, with strong interactions between its structural defects and nanoscale magnetic inhomogenities, despite its deceptively simple cubic rock-salt average structure.\cite{Battle_1979,Wilkinson_1984} W\"ustite has a significant effect on the chemistry of the Earth's lower mantle and is almost always iron deficient, leading to a fraction of the iron oxidising to the trivalent state.\cite{McDonough_2003,Fischer_2011} These cations have been shown to occupy tetrahedral interstitial sites (T) and have a significant effect on the physical properties of w\"ustite, including how it transmits seismic waves.\cite{Koch_1969,Hazen_1984,Zhang_2000,Lin_2003,Zhang_2012} They have also been shown to influence its magnetic structure.\cite{Battle_1979,Wilkinson_1984}

The arrangements of the interstitial iron cations is complex and has been debated for over fifty years.\cite{Koch_1969,Cheetham_1971,Catlow_1975,Andersson_1977,Radler_1990,Welberry_1995,Welberry_1997}  They have been shown to be surrounded by four octahedral vacancies (V) but how individual V$_4$T units connect to each other, as well as their size and distribution, remains unclear. The two most commonly accepted models of these defect clusters have V$_4$T units connected in an edge-sharing or corner-sharing, Koch-Cohen-like, fashion.\cite{Koch_1969,Catlow_1975, Andersson_1977} These differ in nearest-neighbour interstitial iron atoms being related to each other along $\langle100\rangle$ or $\langle110\rangle$ vectors (see Fig.~1). Alternatively the V$_4$T units could be connected  by their corners along $\langle111\rangle$ axes, forming spinel-like clusters similar to the structure of magnetite, Fe$_3$O$_4$,\cite{Andersson_1977} one of the products of w\"ustite disproportionation.\cite{Hazen_1984} 

\begin{figure}
\begin{center}
\includegraphics[width=8.3cm]{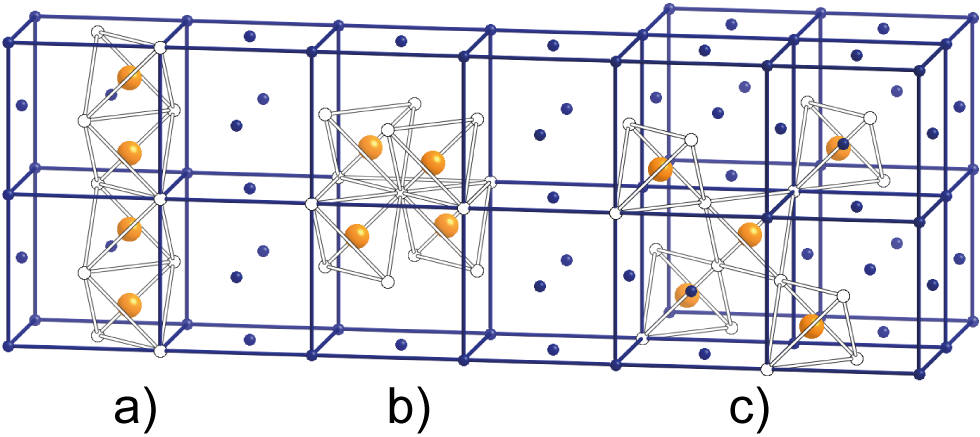}
\end{center}
\caption{\label{fig1} W\"ustite V$_4$T defect units in  a) a $\langle100\rangle$ edge-sharing cluster, b) a $\langle110\rangle$ Koch-Cohen cluster and c) a $\langle111\rangle$ spinel-like cluster. The octahedral iron, tetrahedral iron and octahedral vacancies are dark blue, tangerine and white, respectively}
\end{figure}

Although the defects significantly effect the magnetic properties of w\"ustite the magnetic structure of the defect clusters remains unclear.\cite{Koch_1967,Battle_1979,Wilkinson_1984} Fe$_{1-x}$O orders antiferromagnetically at about 200\,K into a rhombohedral $R\bar{3}$  phase in which the bulk Fe$^{2+}$ moments are said to be ordered ferromagnetically within (111) planes of the parent cubic lattice.\cite{Shull_1951,Battle_1979} The spins are thought to lie parallel to the [111] direction and neighbouring planes are coupled antiferromagnetically. The magnetic moment of the average structure is unusually low, about two-thirds of the expected value, and the tetrahedral iron centres do not exhibit long range magnetic order.\cite{Battle_1979} These observations have been attributed to the iron spins in a tetrahedral cluster lying along a randomly chosen direction in the (111) planes, along with the octahedral irons closest to them due to spin coupling between these octahedral and the tetrahedral cations.\cite{Battle_1979} However reasonable this model might be there is little direct evidence to support it. 

Understanding the details of the local structure of w\"ustite would offer greater insight into the coupling of its nano-sized defect clusters to bulk magnetic ordering in a way that may prove useful in studying a variety of conceptually-related functional materials. Reverse Monte Carlo (RMC) refinements of neutron diffraction total scattering data have recently been shown to be capable of probing the local atomic and magnetic structure of complex materials.\cite{Goodwin_2006,Paddison_2012,Cliffe_2013} In this work we utilised this approach to elucidate the details of the defect and magnetic architecture of w\"ustite. After a brief description of the experimental conditions and the analytical approach employed we discuss the Monte Carlo simulations used to set up the initial models for the RMC refinements. This is followed by analysis of the nuclear structure of w\"ustite, which allows insight into the particular direction along which V$_4$T units connect in order to make larger clusters. Finally the magnetic structure of  w\"ustite will be examined, discussing the magnetic structure of the bulk iron and that of the interstitial cations, as well as the interactions between these two components. We find that the magnetic structure of the bulk iron is more complex than previously thought and that our refinements are consistent with the formation of corner-sharing, Koch-Cohen-like, defect clusters, which in turn support weak ferromagnetic coupling.

\section{Experimental Methods}\label{Experimental Methods}

An 8\,g sample of w\"ustite was made by heating a finely ground mixture of Fe and Fe$_2$O$_3$ in a evacuated sealed quartz vessel at 1173\,K overnight, after which it was quenched in liquid nitrogen.\cite{Battle_1979} The stoichiometry was subsequently determined thermogravometrically to be Fe$_{0.9024(14)}$O. High quality neutron diffraction data suitable for total scattering analysis were collected on the GEM diffractometer at ISIS,\cite{Hannon_2005} at temperatures of 10\,K and 300\,K, chosen to lie well above and below the antiferromagnetic ordering temperature. Data were collected over the range 0.3 $<$ Q $<$ 50\,\AA$^{-1}$. The data were initially fitted by Rietveld refinements in the program GSAS\cite{Larson_1994} using cubic models, since the rhombohedral splitting of the magnetically ordered phase was too small to be observed. Lattice parameters of 4.28989(11)\,\AA\ and 4.2867(4)\,\AA\ were obtained from the 300\,K and 10\,K data, respectively. The displacement parameters of the octahedral and tetrahedral cations were constrained to be equal and their occupancies refined, in a manner that ensured they were consistent with the sample stoichiometry, to 0.844(2) and 0.029(1) against the 300\,K data, respectively. It should be noted that the tetrahedral site has twice the multiplicity of the octahedral site and that the refined values are within error of those obtained from refinements against the 10\,K patterns.

The total scattering data, \emph{F}(\emph{Q}) and \emph{D}(\emph{r}), were then normalised using the program Gudrun\cite{Soper_2011} and models refined using the RMCProfile\cite{Tucker_2007} suite of programs, using a similar methodology to that employed in an earlier study of MnO.\cite{Goodwin_2006} Atomistic models based on $16\times16\times16$ supercells of the cubic unit cell were used with lattice parameters and cation occupancies fixed to the values obtained from the Rietveld refinements. These models contained 13762 octahedral iron (Fe$_{\textrm{oct}}$), 1016 tetrahedral iron (Fe$_{\textrm{tet}}$) and 16384 oxygen atoms; further details are described in Section III.A. It should be noted that while all refined models are metrically cubic RMCProfile\cite{Tucker_2007} does not apply any symmetry elements so the models are allowed to have \emph{P}1 symmetry. For the 300\,K data the paramagnetic contribution to the reciprocal space \emph{F}(\emph{Q}) was modelled by assigning to each cation a spin with random orientation, thereby achieving a paramagnetic arrangement. The corresponding magnetic scattering function was subtracted from the \emph{F}(\emph{Q}) prior to Fourier transform to the real space distribution function, expressed as \emph{D}(\emph{r}). Refinements were carried out against the corrected \emph{D}(\emph{r}) alongside the \emph{F}(\emph{Q}) and back-scattering bank Bragg data.\cite{Keen_2001} RMC refinements against the 300\,K data were performed, in which only atomic positions were allowed to change, by moving randomly-selected atoms by random fractions of a maximum move size in order to minimise the cost function
\begin{equation}
\chi^2_{\mathrm{RMC}} = \sum_m \chi^2_{m}
\end{equation}

 \noindent where the $\chi_m^2$ corresponds to the various data sets being refined. The individual $\chi^2$ functions being minimised were, where $t$ is the neutron time-of-flight:

\begin{equation}
\chi^2_{F(Q)} = \sum_j [F_{\mathrm{calc}}(Q_j) - F_{\mathrm{exp}}(Q_j)]^2\sigma ^{-2}_{F(Q)},
\end{equation}
\begin{equation}
\chi^2_{D(r)} = \sum_j [D_{\mathrm{calc}}(r_j)-D_{\mathrm{exp}}(r_j)]^2\sigma ^{-2}_{D(r)},
\end{equation}
\begin{equation}
\chi^2_{\mathrm{profile}} = \sum_j [I^{\mathrm{calc}}_{\mathrm{profile}}(t_j) - I^{\mathrm{exp}}_{\mathrm{profile}}(t_j)]^2\sigma ^{-2}_\mathrm{profile}
\end{equation}

The Monte Carlo algorithum avoids local minima by accepting individual moves that degrade the fit to the data with a probability that is inversely proportional to how much worse they make the fit. Maximum move sizes of 0.0535 and 0.1000\,\AA\ were used for iron and oxygen atoms, respectively. The weighting, $\sigma$, of each dataset was selected such that the final $\chi^{2}$ values of each were similar and such that between 20 and 50\,\% of attempted moves were accepted. It should be noted that the lowest region of the \emph{D}(\emph{r}) data ($0<r\leq6$\,\AA) was weighted twice as heavily as the rest of this dataset: this range included the first three nearest neighbour interactions of all types of atom pairs, which was considered to be of most interest to this part of the study.

The coherent magnetic scattering present within the 10\,K total scattering data prevents direct Fourier transform of  \emph{F}(\emph{Q}) to a meaningful nuclear \emph{D}(\emph{r}) function. It is, however, possible to calculate the magnetic component of the \emph{F}(\emph{Q}) directly from the atomic positions and their spin orientations in a model.\cite{Blech_1964,Goodwin_2006} This is then added to the nuclear \emph{F}(\emph{Q}), obtained from a Fourier transform of the radial distribution function to calculate the total \emph{F}(\emph{Q}). Consequently our 10\,K refinements were carried out using the \emph{F}(\emph{Q}) and Bragg profile functions only. Refinements allowed atomic positions and cation spin orientations to vary for alternating periods, using similar criteria for weighting data as described above. These refinements started from the equilibrated 300\,K models, including the random Fe spin orientations used to model paramagnetic scattering in which the magnitude for all the octahedral and tetrahedral cations were set to 4 and 5\,${\mu}_\mathrm{B}$, respectively.\cite{Hope_1982,Wilkinson_1984} The Bragg profile from the 24-45$^{\circ}$ bank was used when refining the spin orientation, as this included all the reflections with significant magnetic contribution. In all refinements a closest approach of 1\,\AA\ was applied between all atoms and, based on the observed peak in the \emph{D}(\emph{r}), the minimum iron-oxygen nearest neighbour bond distance was restrained to 1.70\,\AA\ and 1.75\,\AA\ for 10 K and 300 K refinements, respectively. Maximum Fe--O distances of 2.1\,\AA\ and 2.45\,\AA\ for the Fe$_{\textrm{tet}}$ and Fe$_{\textrm{oct}}$ iron were used.\cite{Goodwin_2005}

\section{Results and Discussion}

\subsection{Initial Monte Carlo Generated Models}

The initial configurations used for the RMC refinements were generated by direct Monte Carlo simulations where the positions of the cations and vacancies on both the octahedral and tetrahedral sites were allowed to switch in-order to minimise the sum of an energy term $(4-V)^2$ (where $V$ is the number of octahedral vacancies in the nearest-neighbour shell of a tetrahedral cation). This was done to ensure these models contained V$_4$T units. Additionally half of the models generated included a term to penalise the formation of free octahedral vacancies that did not have a tetrahedral cation in their first coordination sphere to ensure the regions away from the interstitial clusters contained few vacancies. Finally energy terms were also included in some models to encourage the V$_4$T units to connect into clusters along each of the three possible vectors ($\langle100\rangle$, $\langle110\rangle$ and $\langle111\rangle$), while penalising connectivity along the other two. These combinations led to eight different models and in each case RMC refinements were carried out using four distinct configurations to ensure statistical adequacy. The weightings of the various energy terms were modified for each configuration to ensure that at least 98\,\%\ of the tetrahedral interstitial cations were in complete V$_4$T units. Where appropriate, a maximum of 2\,\%\ of octahedral vacancies were allowed to be free and the connectivity of the V$_4$T along the preferred `growth' directions were maximised. 

Comparison of these initial models revealed several significant differences. Firstly those models generated with a term to penalise the presence of free octahedral vacancies had two to three hundred defect clusters in them. This is three to ten times more than the number found in models without such a term, where the defects are typically larger. This difference was particularly stark for models generated with a preferred `growth vector' of $\langle110\rangle$ and $\langle111\rangle$, where the absence of such a penalty results in models with several clusters containing more than one hundred interstitial cations. About 90\,\%\ of the defect clusters in those models with octahedral vacancies concentrated around them have less than ten interstitial cations  and 98\,\%\ of clusters have less than forty tetrahedral cations, except from those with a preferred `growth vector' along the $\langle111\rangle$ vectors have up to 4\,\%\ with more than forty tetrahedral cations. In contrast the models generated without an energy term concentrating the octahedral vacancies near the defect clusters have at least 20\,\%\ of their clusters with more than ten interstitial cations and this increases up to an average of 60(4)\,\%\ and 72(4)\,\%\ for models with V$_4$T units connected along the $\langle110\rangle$ and $\langle111\rangle$ vectors. The larger cluster size of these $\langle110\rangle$ and $\langle111\rangle$ models is emphasised by having an average of 29(14)\,\%\ and 27(2)\,\%\ of their clusters with more than 40 Fe$_{\textrm{tet}}$ centres, although less than  10\,\%\  of  the clusters  in the other models are of this size. The smaller clusters in the models with octahedral vacancies concentrated near the interstitial cations also result in a more uniform distribution of clusters. In all cases, however, there are clearly regions with and without tetrahedral defects, reminiscent of the paracrystalline models previously suggested by Welberry and Christy.\cite{Welberry_1995,Welberry_1997}

\subsection{Local Nuclear Structure}

RMC refinement of the various models against the 300\,K total scattering data resulted in  good fits in all cases (see Fig.~2), although those with the vast majority of octahedral vacancies near the defect clusters gave a better fit (see Table~I). Additionally it appears that the models with V$_4$T units that link into clusters along the $\langle110\rangle$ directions give the best fit to data. This is true both overall and for each type of data fitted, although the difference is predictably subtle. Despite this subtle difference this is our preferred solution and is supported by previous careful single crystal X-ray studies by Koch and Cohen\cite{Koch_1969} and Welberry and Christy\cite{Welberry_1997}. It is an important finding given the precise nature of these defect clusters have been explored for the last fifty years via a variety of techniques without obtaining a definitive model of their structure.\cite{Koch_1969,Cheetham_1971,Catlow_1975,Andersson_1977,Wilkinson_1984,Radler_1990,Welberry_1995,Welberry_1997} In general the crystallographic and magnetic structural details and bonding environments derived from all models are very similar and therefore only the results from those refinements where octahedral vacancies are concentrated near the interstitial clusters will be subsequently discussed. 

\begin{table}
\caption{Statistical measures of fit, including standard deviations between models of the same type, to 300\,K neutron total scattering data. Where applicable the preferred `growth vector' for V$_4$T units is given and model types I and II are those generated without and with an energy term favouring octahedral vacancies concentrating near the defect clusters.}
\begin{tabular}{lllll}\hline

Model & Overall $\chi^{2}$ & D(r) $\chi^{2}$ & F(Q) $\chi^{2}$ & Bragg $\chi^{2}$ \\ \hline
V$_4$T I & 173.4(8) & 185.4(1.6) & 265(3) & 185.4(9) \\
V$_4$T $\langle100\rangle$ I & 173.7(8) & 182.2(9) & 267.2(1.5) & 184.8(1.6) \\
V$_4$T $\langle110\rangle$ I & 171.6(8) & 199.8(1.2) & 258(4) & 181.2(1.8) \\
V$_4$T $\langle111\rangle$ I & 172.9(4) & 209.2(1.6) & 261.4(4) & 179.4(9) \\
V$_4$T II  & 171.3(4) & 171.6(1.1) & 263.5(9) & 183.6(8) \\
V$_4$T $\langle100\rangle$ II & 171.0(6) & 173.1(1.1) & 263.6(1.2) & 182.0(9) \\
V$_4$T $\langle110\rangle$ II & 168.3(8) & 169.6(1.1) & 259(2) & 179.0(8) \\
V$_4$T $\langle111\rangle$ II & 169.9(1.2) & 177.5(1.5) & 261(4) & 179.8(8) \\ \hline

\end{tabular}
\end{table}

\begin{figure}
\begin{center}
\includegraphics[width=8.3cm]{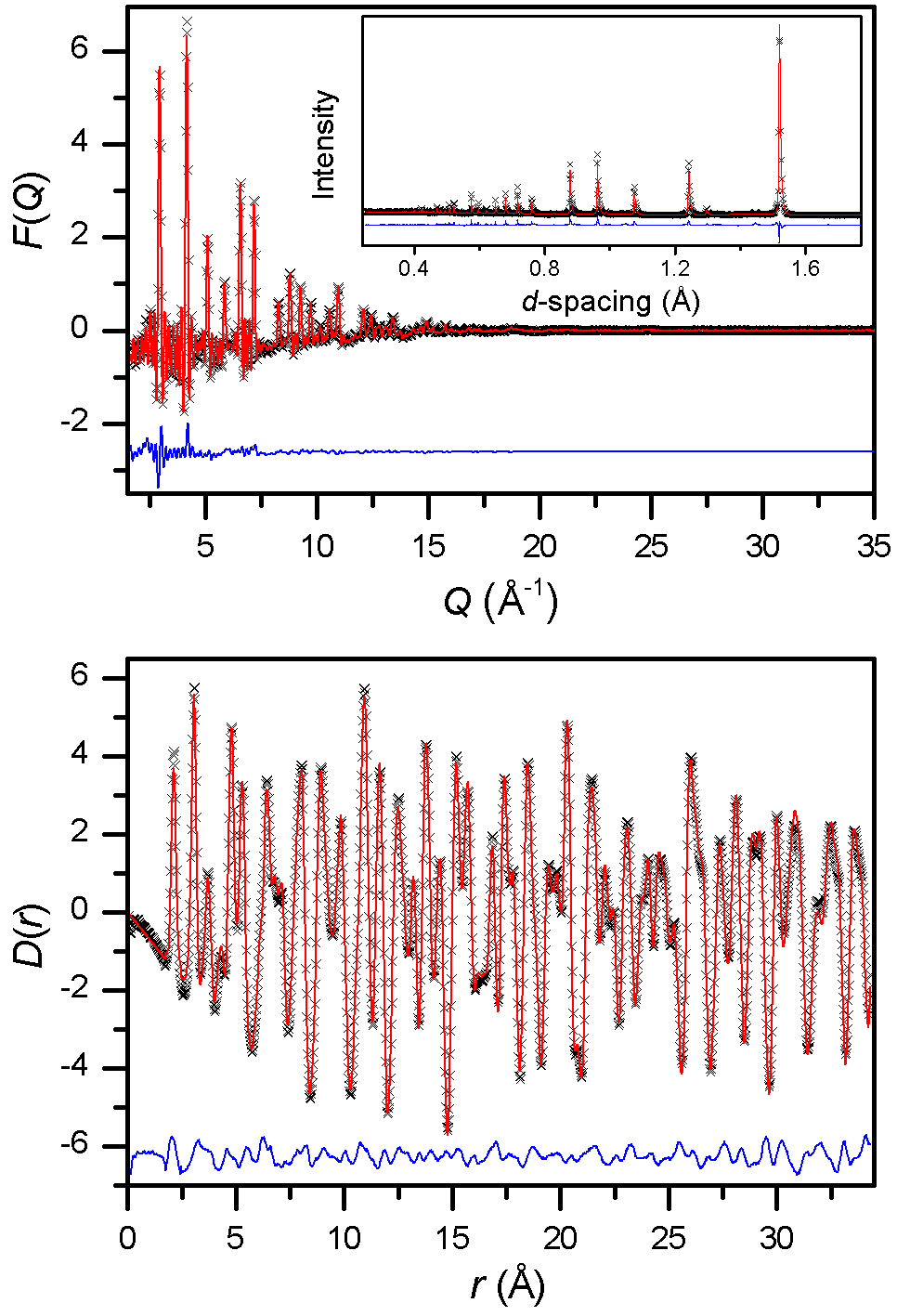}
\end{center}
\caption{\label{fig2} Fits to the 300\,K \emph{F}(\emph{Q}) (top), \emph{D}(\emph{r}) (bottom) and Bragg diffraction data (top inset). The experimental fits are represented by the crosses, fits by red lines and difference plots are in blue.}
\end{figure}

The refined Fe$_{\textrm{oct}}$--O bond distances are very close to the expected values from the average structure: for the bulk iron centres this averages 2.1476(3)--2.1485(3)\,\AA\  (the quoted error here and elsewhere for values obtained from the RMC refinements unless noted otherwise, is the standard error in the mean) across the four different `growth vector' models (see Fig.~3a for a typical octahedral environment). These distances, however, get slightly shorter for Fe$_{\textrm{oct}}$ centres closer to the tetrahedral defect clusters. Those Fe$_{\textrm{oct}}$ atoms with more than two tetrahedral second nearest-neighbours have average bond distances of 2.1343(11)--2.1452(11)\,\AA\ across the four different types of `growth vector' models, with the $\langle110\rangle$ and $\langle100\rangle$ models having the shortest and longest average bond distance, respectively. This change is best reflected by an increase in the average bond valencies from 2.0285(13)--2.0342(13) for the bulk iron to 2.104(4)--2.174(6) for those Fe$_{\textrm{oct}}$ atoms with more than two tetrahedral second nearest-neighbours.\cite{Brown_1985,Brese_1991} The bond angles of the Fe$^{2+}$ centres closer to the defect clusters are also more distorted from an ideal octahedral environment. 

\begin{figure}
\begin{center}
\includegraphics[width=8.3cm]{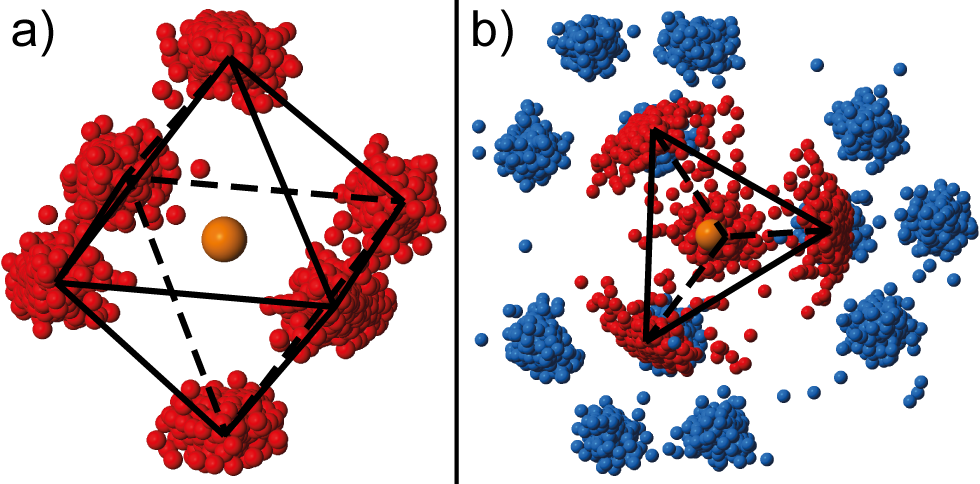}
\end{center}
\caption{\label{fig3}  Typical coordination environment distributions of a) bulk Fe$_{\textrm{oct}}$ cations and b) Fe$_{\textrm{tet}}$ centres with second nearest neighbour Fe$_{\textrm{oct}}$ positions shown. The central iron and oxygen atoms are tangerine and red with the Fe$_{\textrm{oct}}$ in b) in dark blue.}
\end{figure}

The Fe$_{\textrm{tet}}$ atoms appear on a local scale to adopt a regular tetrahedral geometry, with average  Fe$_{\textrm{tet}}$--O bond distances of between 1.9406(7)\,\AA\ and 1.9470(7)\,\AA\, across the four different `growth vector' models and average bond angles of 108.86(6)--109.02(5)$^{\circ}$, with standard deviations of between 8 and 10$^{\circ}$ (see Fig.~3b for typical tetrahedral environment). The bond distance is longer than the value of 1.85750(4)\,\AA\ refined from the average structure, which reduces the average bond valencies of the tetrahedral cation from 2.86 to 2.315(4)--2.353(4).\cite{Brown_1985,Brese_1991} This is somewhat surprising since previous M\"{o}ssbauer studies have indicated that the Fe$_{\textrm{tet}}$ cations are trivalent.\cite{Hope_1982,Wilkinson_1984} Consistent with previous studies, however, it appears that both the tetrahedral iron and the octahedral iron closer to the defect clusters have an average higher charge than the bulk.\cite{Hope_1982,Wilkinson_1984}

That the tetrahedral cations may not be purely trivalent at high temperatures demands an increase of the negative charge on the defect cluster regions. This is caused by having more octahedral vacancies than tetrahedral interstitials in these regions, as required by an overall ratio of 2.7 vacancies per interstitial cation. This may be compensated for somewhat by the shortening of the mean Fe$_{\textrm{tet}}$--Fe$_{\textrm{oct}}$ second and third nearest-neighbour distances. These average 3.4629(9)--3.4783(9)\,\AA\ and 4.6074(8)--4.6183(7)\,\AA\ across the four different types of `growth vector' models --- about 0.09\,\AA\ and 0.06\,\AA\ shorter than the values obtained from the average structure, indicating that the Fe$_{\textrm{oct}}$ cations nearest the clusters are drawn towards them. 

Examining the oxygen bond valencies reveals that those oxygens that bond to either only Fe$_{\textrm{oct}}$ or a mixture of Fe$_{\textrm{oct}}$ and Fe$_{\textrm{tet}}$ centre have reasonable average bond valencies, of 1.8493(10)--1.8713(10) and 1.826(3)--1.942(4), respectively, across the four different `growth vector' models (see Fig.~4a for bond valency distribution).\cite{Brown_1985,Brese_1991} The majority of the oxygens that bond to octahedral cations are six coordinate while those that bond to a mix of cation types are overwhelmingly four coordinate, mostly bonding to three octahedral iron and one tetrahedral cation. Conversely the bond valencies and coordination numbers of oxygen anions that only bond to the tetrahedral cations are highly dependent on the particular direction in which the V$_4$T units connect. Only the $\langle110\rangle$ and $\langle111\rangle$ models have bond valencies close to the expected value of two (2.11(3) and 1.93(4), respectively) and relatively uniform coordination numbers, with the majority being four coordinate and about a quarter in each being three coordinate.\cite{Brown_1985,Brese_1991} The distribution of bond valencies in the $\langle110\rangle$ models are especially regular (see Fig.~4b) and, while only about 1\,\%\ of oxygen anions bond exclusively to tetrahedral cations, this is clear crystal-chemical support in favour of this model. A slice through a typical refined $\langle110\rangle$ model, in which most Fe$_{\textrm{oct}}$ vacancies are concentrated near the defect clusters, is shown in Fig. 5.  

\begin{figure}
\begin{center}
\includegraphics[width=8.3cm]{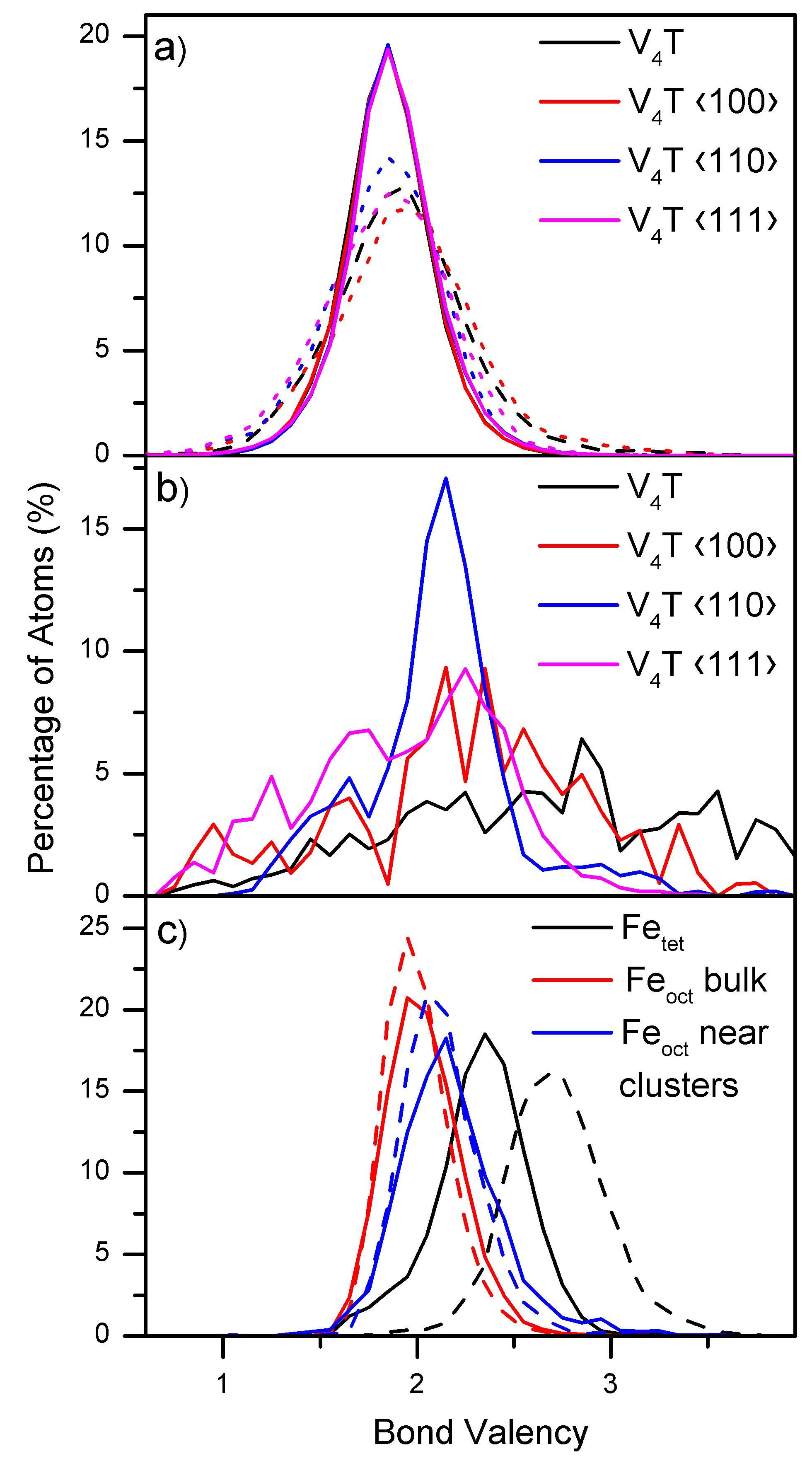}
\end{center}
\caption{\label{fig4} Average bond valency distribution of a) oxygen atoms bonded to only octahedral cations (continuous lines) and both types of Fe (dashed line) and  b) oxygen anions bonded only to Fe$_{\textrm{tet}}$ centres. c) indicates the average bond valency distribution, across the $\langle110\rangle$ models,  of Fe$_{\textrm{tet}}$, bulk Fe$_{\textrm{oct}}$ and Fe$_{\textrm{oct}}$ cations with more than two Fe$_{\textrm{tet}}$ second nearest neighbours at 300\,K (continuous lines) and 10\,K (dashed line). Bond valences are binned in a range of 0.1.}
\end{figure}

\begin{figure}
\begin{center}
\includegraphics[width=8.3cm]{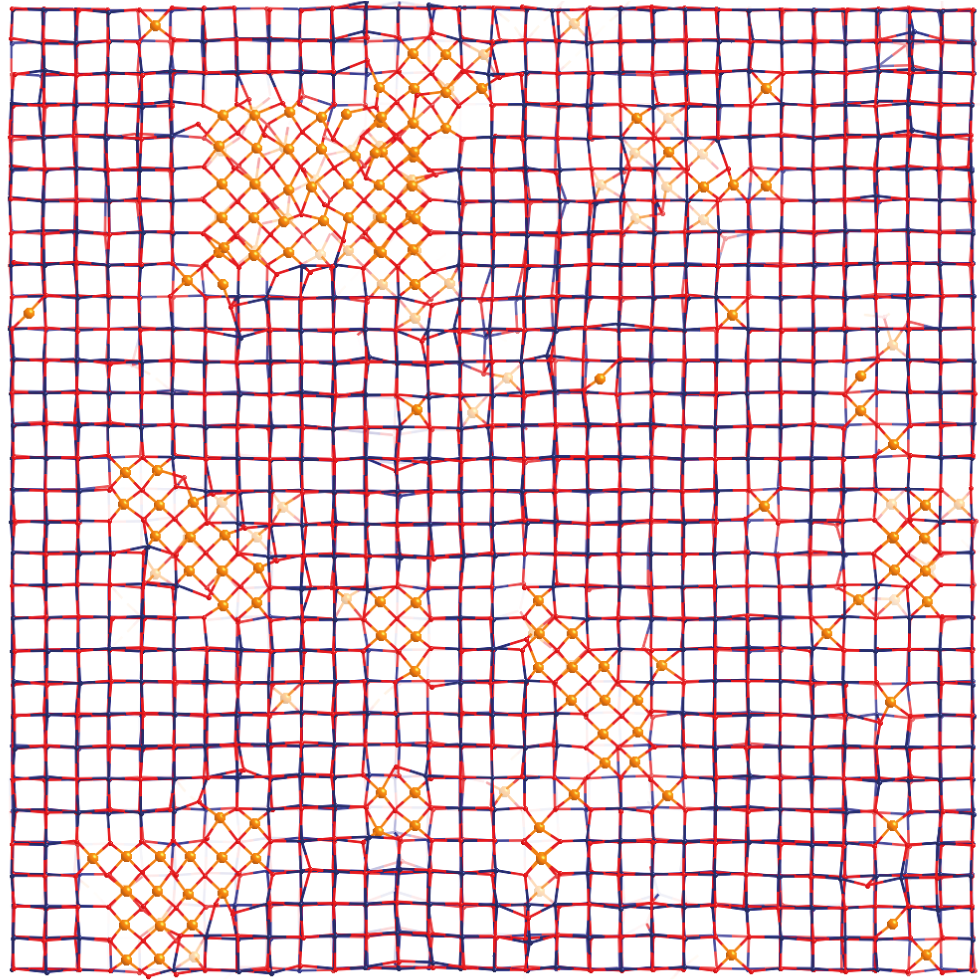}
\end{center}
\caption{\label{fig5} A slice through a typical $\langle110\rangle$ model refined from total scattering data, highlighting the clusters of interstitial cations (shown in tangerine). The view is down the $c$-axis and Fe$_{\textrm{oct}}$ and oxygen atoms are shown in dark blue and red, respectively.}
\end{figure}

Refinements against the 10\,K data indicated that, while retaining the same coordination geometry, the Fe$_{\textrm{tet}}$--O bond distances shorten significantly with averages of between 1.8893(8)\,\AA\ and 1.8914(8)\,\AA\ across the different types of models. This increases the mean tetrahedral bond valencies to 2.704(4)--2.720(4) (see Figure~4c for bond valence distribution).\cite{Brown_1985,Brese_1991} Significantly the average Fe$_{\textrm{tet}}$--Fe$_{\textrm{oct}}$ second and third nearest neighbour distances also increase by about 0.03\,\AA, which is coupled with a small decrease in Fe$_{\textrm{oct}}$ cation bond valencies, particularly in the case of the Fe$_{\textrm{oct}}$ cations closest to the interstitial clusters. This increase in the apparent charge on the Fe$_{\textrm{tet}}$ cations while the Fe$_{\textrm{oct}}$ centres close to the defect clusters move slightly away from them may suggest a change in the charge distribution between 300\,K and 10\,K. This is consistent with the work of Hope \emph{et al.}\cite{Hope_1982} and Wilkinson \emph{et al.}\cite{Wilkinson_1984} who suggested, based on M\"{o}ssbauer spectroscopy, that there is charge transfer from the Fe$_{\textrm{tet}}$ cations to the Fe$_{\textrm{oct}}$ centres at higher temperature near the clusters, which ceases at lower temperature.

\subsection{Local Magnetic Structure}

Initial examination of the average orientation and correlations of the spins of the bulk octahedral atoms at 10\,K data suggests they are broadly consistent with the magnetic structure previously proposed by Battle and Cheetham \cite{Battle_1979} (see Fig.~6a and 7a). Careful examination of the average spin orientation of each of the 32 Fe$_{\textrm{oct}}$ atoms in a double unit cell, however, reveals a more complicated structure. While it appears that the previous description of ferromagnetically coupled (111) planes coupled antiferromagnetically to each other is broadly correct, the spins are not all oriented along a $\langle111\rangle$  vector. The spin orientation of the Fe$_{\textrm{oct}}$ atoms within the four groups of atoms generated by the face-centering operation are consistent with an antiferromagnetic model but the orientation of the spins between these four different groups is different (see Fig.~6b-e). Our refinements do not appear to be sensitive to their precise orientation, which varies somewhat between different refinements. The simplest model consistent with the RMC refinements, however, would have the spins of one of these groups of atoms oriented directly along the $\langle111\rangle$ direction while each of the other three groups of atoms have their spins orientation slightly towards, or away from, one of the three $\langle100\rangle$ directions. 

\begin{figure}
\begin{center}
\includegraphics[width=8.3cm]{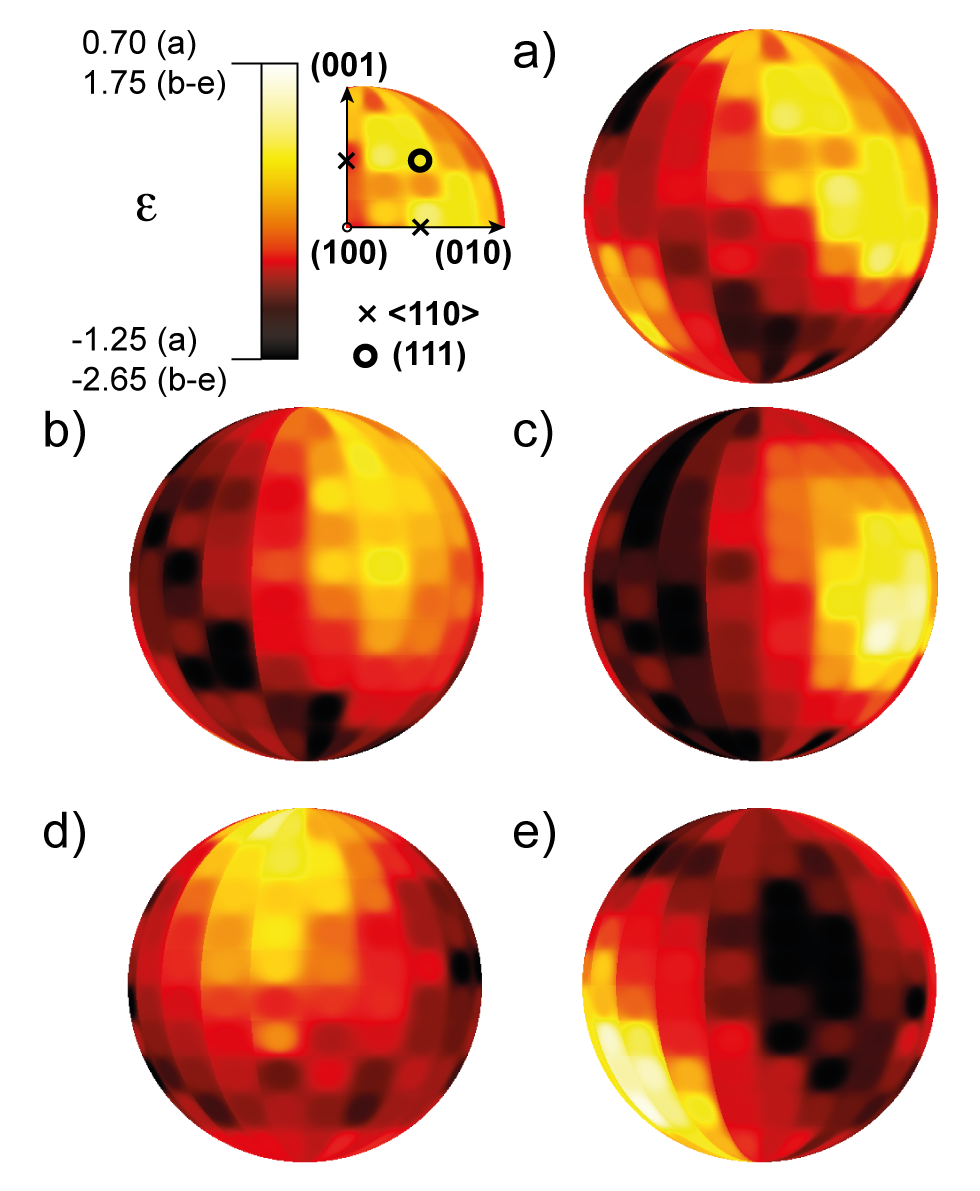}
\end{center}
\caption{\label{fig6}  Orientation of the Fe$_{\textrm{oct}}$ spins in a typical model of w\"ustite refined by RMC illustrating a) all the bulk Fe$_{\textrm{oct}}$ spins  and b-e) the four groups of octahedral cations generated by the face-centring operation. The relative spin density, $\ensuremath{\epsilon\ensuremath{(\theta,\phi)}}$ is defined as $\ensuremath\epsilon(\theta,\phi)=\ln\left[\frac{n(\theta,\phi)}{N{\rm d}(\cos\theta)\,{\rm d}\phi}\right]$, where $\ensuremath{n(\theta,\phi)}$ is the number of spins with orientations within the range $\ensuremath{{\rm d}(\cos\theta),\mathrm{d}\phi}$.}
\end{figure}

\begin{figure}
\begin{center}
\includegraphics[width=8.3cm]{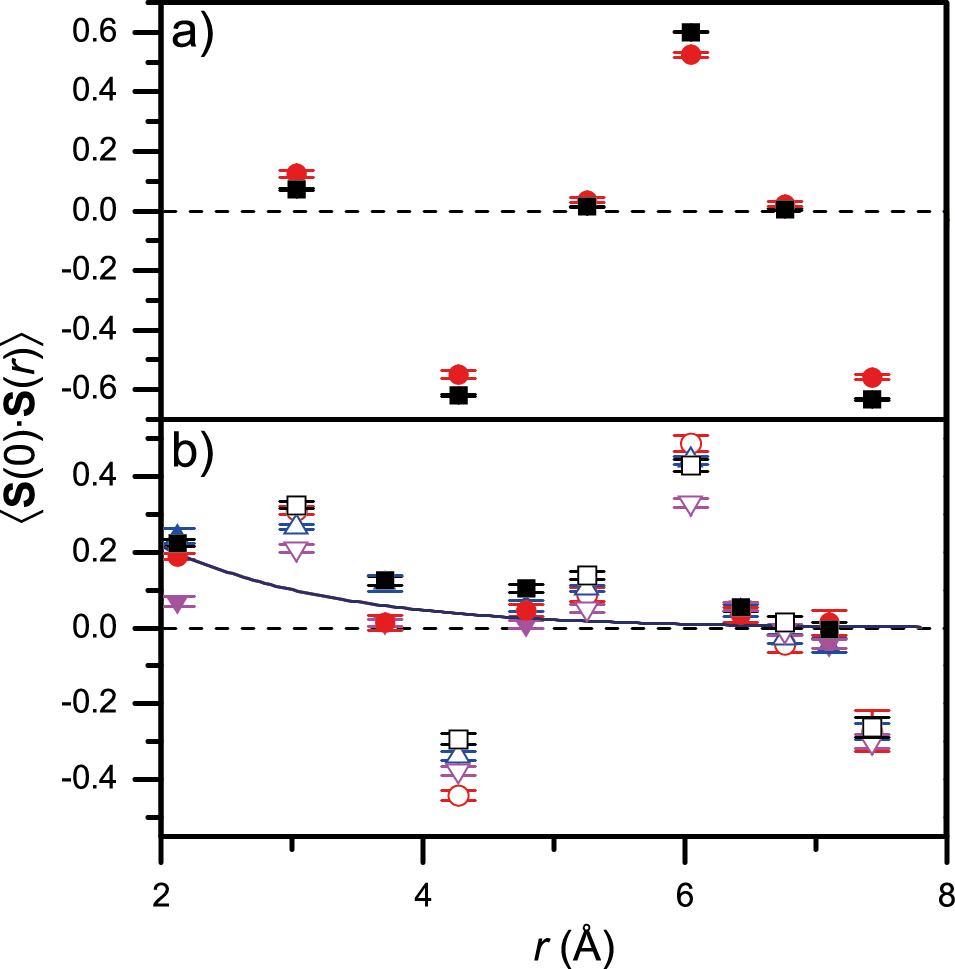}
\end{center}
\caption{\label{fig7}  a) Average $\langle110\rangle$ model Fe$_{\textrm{oct}}$--Fe$_{\textrm{oct}}$ spin correlations of bulk iron (square markers) and those with more than two tetrahedral nearest neighbours (circle markers) and b) average Fe$_{\textrm{tet}}$--Fe$_{\textrm{tet}}$ correlations from all models with V$_4$T units connected without a preferred direction (square markers) and those along $\langle100\rangle$ (circle markers), $\langle110\rangle$ (up triangles) and $\langle111\rangle$ (down triangles) vectors. In b) the line is the correlation length fit to the filled markers, which are those that do not overlap with Fe$_{\textrm{oct}}$--Fe$_{\textrm{oct}}$ spin correlations.} 
\end{figure}

We have carried out Rietveld refinements of such a model and we find that this provides a superior fit to data compared to the previously reported model,\cite{Battle_1979} despite having only one extra parameter (\emph{c.f.} $R_{\textrm{p}}$ and $R_{\textrm{wp}}$ of 4.8\,\%\  and 4.0\,\%\ for a four site model to 5.1\,\%\  and 4.6\,\% for the collinear model). The spin orientations refined such that the $\langle110\rangle$-like component is 1.477(13)\,${\mu}_\mathrm{B}$ while the $\langle100\rangle$-like component was higher, 2.511(10)\,${\mu}_\mathrm{B}$. The magnitude of the spins oriented along the (111) axis were set to have equal magnitude to the other atoms; a moment of 2.913(15)\,${\mu}_\mathrm{B}$ (see Fig.~8 for depiction of this refined structure). This is still only 73\,\%\ of the full moment expected for an Fe$^{2+}$ cation although it is somewhat higher than that refined by Battle and Cheetham\cite{Battle_1979} 

\begin{figure}
\begin{center}
\includegraphics[width=8.3cm]{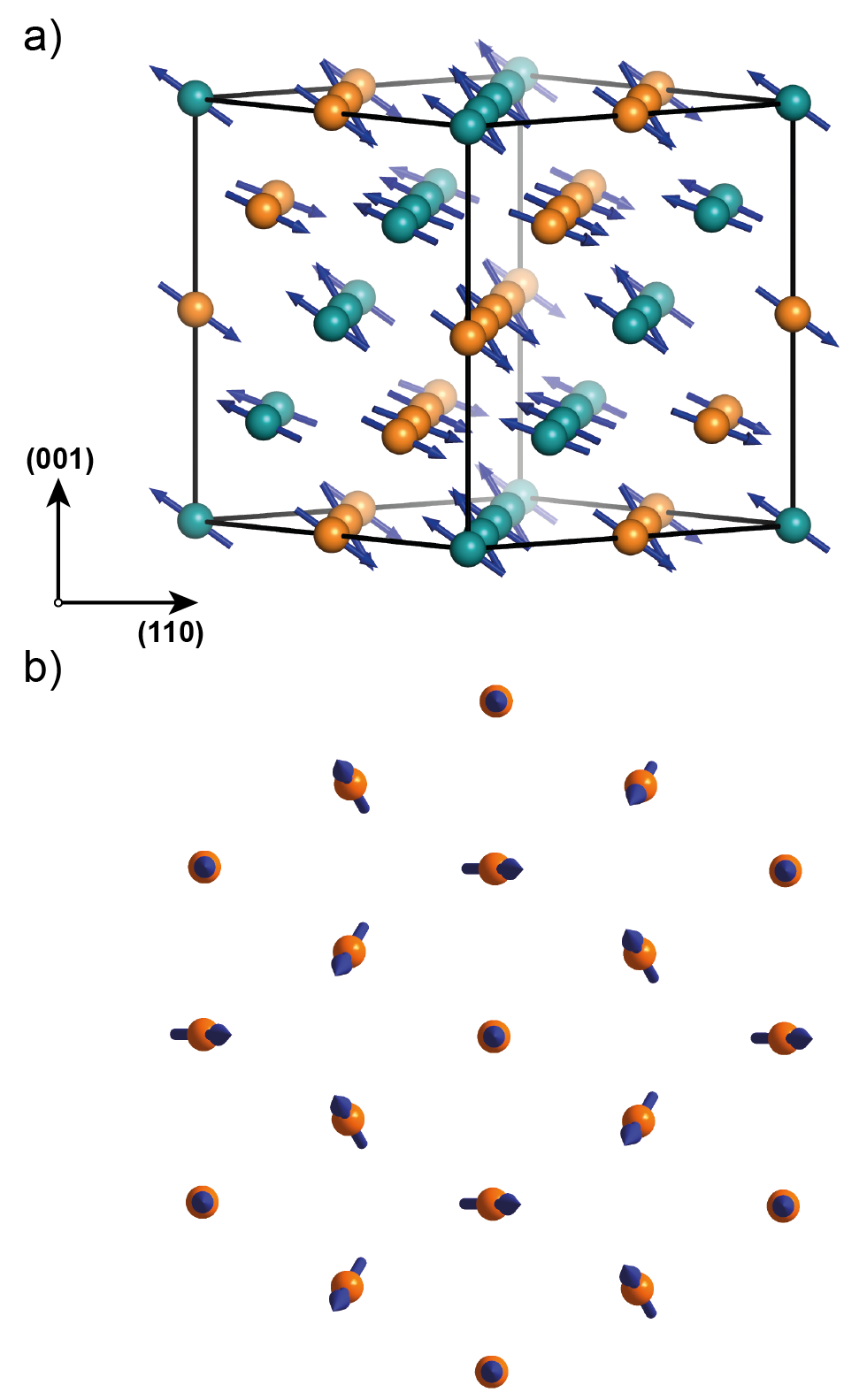}
\end{center}
\caption{\label{fig8} The average bulk magnetic structure of w\"ustite, refined by the Rietveld method, showing a) antiferromagnetically coupled [111] planes of Fe$_{\textrm{oct}}$ atoms, presented in different colours, and b) the orientation of the Fe$_{\textrm{oct}}$  cations within a single ferromagnetically coupled plane.}
\end{figure}

Support for this four site model can also be found in polarised neutron diffraction studies by Wilkinson \emph{et al.}\cite{Wilkinson_1984} which suggested that, even in the inter-cluster bulk region, the average magnetic moments lie at approximately a 30$^{\circ}$ angle from the [111] axis. In our refined model those Fe$_{\textrm{oct}}$ cations whose spins deviate from the [111] axis do so by a higher angle, about 45$^{\circ}$; since one quarter of the spins in our model point along the [111] axis. The total deviation from this direction in our model and that of Wilkinson \emph{et al.}\cite{Wilkinson_1984} is very similar. Overall our model is very reminiscent of a condensed spin-wave, such as that found in the related magnetic structure of MnO.\cite{Goodwin_2006,Goodwin_2007} This view is emphasised when looking at the magnetic structure of a single ferromagnetically coupled (111) plane of w\"ustite (see Fig.~8b). Here it can be seen that any atom is surrounded by six nearest neighbours, which can be broken down into three alternating groups of two atoms, each group having a spin orientation not adopted by the central atom; the second-nearest neighbour Fe$_{\textrm{oct}}$ have the same arrangement offset by 90$^{\circ}$. 

The spin distribution and correlations of the octahedral cations closest to the tetrahedral clusters are similar to the bulk but have a broader distribution. This is in contrast with the work of Battle and Cheetham\cite{Battle_1979} and Wilkinson \emph{et al.}\cite{Wilkinson_1984}, who suggested the moments of the Fe$_{\textrm{oct}}$ close to the defect clusters lie in or close to the (111) plane. Given the previous models were primarily inferred from a lower than expected refined magnetic moment rather than a direct result of fits to experimental data, as is the case in this work, we believe it is more likely that the model in the current study is correct. It is possible, however, that the moments of the Fe$_{\textrm{oct}}$ close to the defects do lie close to the (111) plane, as suggested by these previous studies, and our data is insensitive to this feature.  Fe$_{\textrm{oct}}$ atoms appear to couple in a weakly ferromagnetic fashion to their second nearest Fe$_{\textrm{tet}}$ neighbours, with an average $\langle$Fe$_{\textrm{oct}}(0)\cdot$Fe$_{\textrm{tet}}(r)$$\rangle$ spin correlation of 0.217(7) across the $\langle110\rangle$ models at an average distance of 3.4928(6)\,\AA. It should be noted that this term would be equal to 1 in a completely-ordered ferromagnet at 0\,K and $-1$ in a pure antiferromagent. In contrast the coupling between the spins of Fe$_{\textrm{oct}}$ cations and their third nearest Fe$_{\textrm{tet}}$ neighbours are weakly antiferromagnetic, with an average spin correlation of $-$0.213(6) at a distance of 4.6302(6)\,\AA, although this decreases in magnitude as the octahedral iron get closer to the clusters. The direction between the octahedral cations and their second and third nearest neighbour Fe$_{\textrm{tet}}$ atoms are similar, along the $\langle211\rangle$ and $\langle221\rangle$ vectors, so we suggest the differences in the coupling with these atoms is related to the distance between the cations. It is important to note that the distances at which these interactions occur are distinct from those at which Fe$_{\textrm{oct}}$--Fe$_{\textrm{oct}}$ correlation take place so these results should be free of interference from such coupling.  Over half the octahedral cations have a tetrahedral atom in the third nearest neighbour shell and around a third of the octahedral cations have a tetrahedral cation in their second coordination sphere. Therefore the spin orientation of most octahedral cations will be affected by coupling with tetrahedral neighbours and the resulting deviation of the magnetic structure from that of the defect free regions is the likely cause of the reduction of the average spin moment obtained from the Rietveld refinements.

Examination of the refined spin distribution of the tetrahedral cations indicates that, overall, these have a random distribution and do not favour lying in the (111) plane, as previously hypothesised.\cite{Battle_1979} Unfortunately, the size of the clusters was too small to examine if there are any trends within individual defects. The strongest Fe$_{\textrm{tet}}$--Fe$_{\textrm{tet}}$ coupling apparent in spin correlations plots are at the same distances as strong Fe$_{\textrm{oct}}$--Fe$_{\textrm{oct}}$ correlations are observed (see Fig. 7b). We interpret this as being due to the Fe$_{\textrm{oct}}$ atoms in w\"ustite dominating its magnetic scattering due to their higher concentration. Therefore, the signal from the Fe$_{\textrm{tet}}$ is much weaker and where  Fe$_{\textrm{oct}}$--Fe$_{\textrm{oct}}$ correlations exist the Fe$_{\textrm{tet}}$ spins refine in a way that is consistent with this coupling. Excluding regions effected by such coupling the general trend of Fe$_{\textrm{tet}}$--Fe$_{\textrm{tet}}$ correlations suggests the clusters have weak ferromagnetic interactions. The correlation length has been determined to be 1.31(9)\,\AA, calculated a from fit to the average of the spin correlations functions for distances free from Fe$_{\textrm{oct}}$--Fe$_{\textrm{oct}}$ contamination (see Fig.~7b). The small correlation length suggests the coupling does not extend far beyond nearest neighbour Fe$_{\textrm{tet}}$--Fe$_{\textrm{tet}}$ interactions. 

 Although there are some limitations in regards to the details obtained of the magnetic structure of the interstitial atoms, the greater insight into the magnetic structures of both the defect clusters and the bulk iron achieved in this study shows the clear advantages of the approach used in this work to analyse total scattering data. The first is the sensitivity of total scattering to both local nuclear and magnetic structures, which allows the interactions between these two aspects to be probed. That the RMC method probes total scattering data by refining large atomistic models in a stochastic fashion is also advantageous over other methods for analysing total scattering data, as seen from the refinement of the magnetic structure from a random arrangement of spins in the absence of a preconceived model.\cite{Egami_2012} This approach allowed the non-collinear nature of the bulk Fe$_{\textrm{oct}}$ magnetic structure to be established and then subsequently confirmed by an improved Rietveld refinement of the average structure, when previous studies had overlooked this feature. While the weaker sensitivity of the data to the relatively low proportion of iron in the tetrahedral clusters (about 7\,\%), leads to the correlations of the bulk dominating somewhat where these overlap, this could at least partly be improved through other measurements, such as polarised neutron scattering. Using data obtained at different temperatures was of great advantage in this respect as it allowed the  nuclear structure to be examined free from the influence of magnetic order and then the magnetic interactions to be probed once a better insight into the nuclear architecture had been established.

\section{Conclusion}
Our analysis suggests the structure of w\"ustite is best described as islands of weakly ferromagnetically coupled Fe$_{\textrm{tet}}$ clusters, most likely with intra-cluster connectivity along the $\langle110\rangle$ axes. These are then surrounded by a bulk Fe$_{\textrm{oct}}$ structure, the magnetic architecture of which is a non-collinear variant of type-II antiferromagnetic order. It appears that most octahedral vacancies are found near to the defect clusters. There is also evidence of both charge and spin interactions between the defects and the bulk. There appears to be a change in charge distribution between ambient and low temperature, with the charge on the Fe$_{\textrm{tet}}$ cations increasing at lower temperatures. The Fe$_{\textrm{oct}}$ atoms nearest the clusters have broader spin distributions than the bulk, attributed to coupling to the Fe$_{\textrm{tet}}$ cations. Despite some limitations, particularly with regards to the magnetic structure of the interstitial iron cations, outlined above this work is, to the best of our knowledge, the first total scattering study to analyse the structure of w\"ustite and hence to probe closely the relationship between its cluster defect structure and its magnetic architecture. It is expected that similar approaches to materials in which there is strong coupling between structural defects and their magnetic or charge-ordering structures will provide significant insight into important advanced materials.\cite{Khajetoorians_2012,Orenstein_2000,Mazin_2008,Senn_2012}

\textbf{Acknowledgements}
 The authors would like to acknowledge the funding of this work by the European Research Council (ERC), through its 'Starting Grant' (number 279705), and the Engineering and Physical Sciences Research Council (EPSRC), UK through grants EP/G004528/1 and EP/G004528/2. The authors would also like to thank the Science and Technologies Facilities Council (STFC) for access to the GEM diffractometer at ISIS. P.J.S thanks Callum Young for helpful discussions and his work modifying RMCProfile allowing improved handling of magnetic refinements. A modified version of this article has been accepted for publication in Physical Review B (prb.aps.org)

 \end{document}